\documentclass[article,twocolumn,showpacs,aps]{revtex4-1}

\expandafter\let\csname equation*\endcsname\relax
\expandafter\let\csname endequation*\endcsname\relax

\usepackage{amsmath}
\usepackage{amssymb}
\usepackage{color}
\usepackage{verbatim}
\usepackage{array}
\usepackage{graphicx}
\usepackage{placeins}
\usepackage{epsfig}
\usepackage{bm}
\usepackage[ragged]{sidecap}
\usepackage{dsfont}
\usepackage{ulem}
\usepackage{mathrsfs}
\usepackage{setspace}

\allowdisplaybreaks

\newcommand{\ds}{\displaystyle}

\newcommand{\supercomas}[1]{``#1''}

\newcommand{\opvector}[1]{\hat{\mathbf{#1}}}
\newcommand{\opscalar}[1]{\hat{\textrm #1}}

\newcommand{\dd}{\mathrm{d}}

\newcommand{\bra}[1]{\mathinner{\langle{#1}|}}
\newcommand{\ket}[1]{\mathinner{|{#1}\rangle}}

\newcommand{\braket}[2]{\langle #1|#2\rangle}
\newcommand{\uvector}[1]{\underline{#1}}

\definecolor{bb}{RGB}{0,0,127}
\definecolor{LightRed}{RGB}{255,100,100}
\definecolor{DarkRed}{RGB}{105,0,0}

\newcommand{\comm}[2]{\left[#1,#2\right]}

\newcommand{\abs}[1]{\lvert #1 \rvert}
\newcommand{\bbarOp}{\opvector{b}\hspace{-0.3cm}^-}

\begin{document}
\preprint{\hfill\parbox[b]{0.3\hsize}{ }}

\title{Two-photon processes based on quantum commutators
}

\author{
F. Fratini$^{1,}$\footnote{
\begin{tabular}{ll}
E-mail addresses:& fhb159809@fh-vie.ac.at\\
& fratini.filippo@gmail.com
\end{tabular}},
L. Safari$^2$,
P. Amaro$^{3}$,
J. P. Santos$^{2}$,
}
\affiliation
{
\begin{tabular}{c}
$^1$ University of Applied Sciences BFI Vienna, Wohlmutstra\ss e 22, A-1020 Wien, Austria\\
$^2$ IST Austria, Am Campus 1, A-3400 Klosterneuburg, Austria\\
$^3$ Laborat\'orio de Instrumenta\c{c}\~ao, Engenharia Biom\'edica e F\'isica da Radia\c{c}\~ao
(LIBPhys-UNL),\\Departamento de F\'isica, Faculdade~de~Ci\^{e}ncias~e~Tecnologia,~FCT, \\Universidade Nova de Lisboa,~P-2829-516 Caparica, Portugal.
\end{tabular}
}

\date{\today}

\begin{abstract}
We developed a new method to calculate two-photon processes in quantum mechanics that replaces the infinite summation over the intermediate states by a perturbation expansion. This latter consists of a series of commutators that involve position, momentum and hamiltonian quantum operators. 
We analyzed several single- and many-particle cases for which a closed form solution to the perturbation expansion exists, as well as more complicated cases for which a solution is found by convergence. 
%
%
Throughout the article, Rayleigh and Raman scattering are taken as examples of two-photon processes. 
The present method provides a clear distinction between the Thomson scattering, regarded as classical scattering, and quantum contributions. Such a distinction let us derive general results concerning light scattering.
%
Finally, possible extensions to the developed formalism are discussed.
\end{abstract}


\maketitle

\section{Introduction}

Two-photon processes are of utmost importance for a variety of applications and techniques in chemistry and biology \cite{jacques2013optical, Ces1992, ssw2009, kwh1997}, including two-photon excited fluorescence microscopy \cite{mwe1995, abd1998, lzw2003}, optical imaging  \cite{ewl1997, pcd2009}, three dimensional optical data storage \cite{swe1991, kka2000},  two-photon induced biological caging studies \cite{den1994, wgb2010}, and also analysis of mesoscopic systems \cite{bartsch1992dynamic}. In all the mention techniques, accurate theoretical predictions of both two-photon absorption and light scattering cross-sections are highly demanded for the search of molecules and specimens with the largest cross-sections, and thus with highest contrast. 

The main full quantum mechanical approaches  to calculate  two-photon cross-sections are either third-order polarizabilities \cite{tkb2008, mts2004, owa1971}, or the dispersion theory of Kramers-Heisenberg  \cite{mra2010,th1982}, as well as its relativistic analogous - the S-matrix approach \cite{alb1965, roy1999}. All of these approaches contain a summation over the infinite intermediate states of the target bound system. In case of one-electron atomic systems,  such intermediate-state summation has been evaluated for a variety of second-order processes, leading to high accurate values of cross-sections and emissions rates \cite{gmi1967, fmp1990, lki2004, dad1992, mil1970, sis2011, sfi1998, Pedro2009, god1981}, as well as its dependence upon photon polarizations \cite{saf2012b, sas2015, fan2011, rkl1986} and geometry \cite{saf2012, wvs2017}. 
Nevertheless, in case of many-body systems or complex potentials, such a summation over the infinite intermediate states is often difficult, or impossible to be evaluated accurately \cite{dra1985, lau1980, sch1973, svf2010, vys2016, aff2016}. When considering molecules, this summation is even harder to perform due to either the complexity of obtaining states for complex potentials or the summation requiring a huge amount of vibrational and rotational states for a reliable evaluation, even for harmonic potentials \cite{th1982}.
Because of these reasons, simpler methods - such as the Thomson or the Form Factor (FF) approximations \cite{Roy1993, ssa2015} - are very much used when calculating, for instance, light scattering by many-electron atoms or crystallographic specimens \cite{Giac, ewl1997, pcd2009}, although they are unable to capture quantum mechanical effects that are given by the target bound spectrum, such as quantum interference \cite{amaro2015, amaro2015quantum} or resonance effects \cite{Pedro2009}, among others. 

Here, we propose a new method to calculate two-photon processes that replaces the infinite summation over the intermediate states by a perturbation expansion. This latter consists of a series of commutators that involve position, momentum and hamiltonian quantum operators. Thus, the problem of describing two-photon processes is moved from solving complex Schr\"{o}dinger equations and Green functions - so to find the intermediate states to be summed - to computing a series of commutators. 
We show several cases for which a closed form solution to the perturbation expansion exists, as well as cases for which the solution is found by convergence. Moreover, our analytical method will allow us to make statements in the form of rules that the two-photon process must obey. 

For simplicity and brevity, we shall restrict our analysis to light (Rayleigh or Raman) scattering, which is one of the most interesting two-photon processes due to its interdisciplinarity. In fact, Rayleigh and Raman scattering, besides being the main tools used to analyze molecular specimens in diverse areas of science \cite{Movasaghi2007, RayMed, Giac}, are also the basic processes in quantum communication for upcoming technologies based on light propagation at the single photon level \cite{kimble2008, warburton2008, fratini2016full}. The reader will notice that the formalism here developed is general and can be applied to any two-photon process, such as two-photon decay or two-photon absorption. Finally, conclusions and possible extensions to the developed formalism are discussed at the end of the article.

SI units are used throughout the article, unless differently specified.

\section{Light scattering off bound states}
\label{sec:BasicTheory}
\begin{figure}[b]
\includegraphics[scale=0.4]{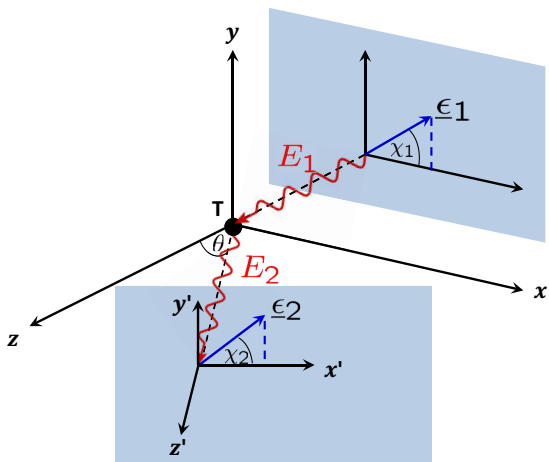}
\caption{
(Color online) The light scattering process. An incident photon with energy $E_1$ and linear polarization vector $\underline{\epsilon}_1$ scatters off a target T. The scattered photon has energy $E_2$ and linear polarization vector $\underline{\epsilon}_2$. $\theta$ is the scattering angle, while $xz$ plane is the scattering plane.
}
\label{fig:geometry}
\end{figure}

We shall work within the dipole approximation. Such an approximation is justified if the light wavelength is much larger than the size of the target. An extension of the present work to higher multipoles is possible, for which details are provided in Sec. \ref{sec:Ext}.

Light scattering is described in non-relativistic quantum mechanics by the Kramer-Heisenberg formula. The (polarization dependent) differential cross section for such a process reads \cite{tulkki1982behaviour}
\begin{equation}
\ds\frac{\dd \sigma^{\uvector\epsilon_1\uvector\epsilon_2}}{\dd \Omega}=
r_e^2\frac{E_2}{E_1}|\mathcal{M}|^2~,
\end{equation}
where $r_e$ is a constant, $E_{1(2)}$ and $\uvector\epsilon_{1(2)}$ are the energy and polarization vector of the incoming (outgoing) photon, respectively. For atomic targets, $r_e$ is equal to the classical electron radius. Considering a target composed by $N$ charged compounds with mass $m$, the scattering amplitude $\mathcal{M}$ is defined as
\begin{equation}
\begin{array}{lcl}
\mathcal{M}&=&\ds 
N
\bra{f}\uvector{\epsilon}_1\cdot \uvector{\epsilon}_2^*\ket{i} -
\frac{1}{m}\left(\mathcal{A}_{12}+\mathcal{A}_{21}\right)~.
\end{array}
\label{eq:Ampl}
\end{equation}
The term $\mathcal{A}_{12}$ reads
\begin{equation}
\begin{array}{lcl}
\mathcal{A}_{12}&=&\ds\sum_\nu\sum_{j, t=1}^N
\frac{
\bra{f}\opvector{p}_{j}\cdot \uvector{\epsilon}_2^*\ket{\nu}\bra{\nu}\opvector{p}_t\cdot\uvector{\epsilon}_1\ket{i}
}{
E_\nu-E_i-E_1
}~,
%
\end{array}
\label{eq:Aampl}
\end{equation}
while $\mathcal{A}_{21}$ is obtained from $\mathcal{A}_{12}$ by replacing $E_1\to-E_2$ and $\uvector \epsilon_1\leftrightarrow \uvector \epsilon_2$.
$\opvector p$ is the momentum operator. We denote by $E_{i,f}$ the energies of initial and final states of the target. On the other hand, $E_\nu$ are the energies of the intermediate states to be summed for the computation of the amplitude. As in \cite{saf2012b, sas2015}, we shall consider light scattering as depicted in Fig. \ref{fig:geometry}. Without restriction of generality and unless differently specified, we consider linearly polarized photons, with $\chi_{1(2)}$ being the azimuthal angle that defines the incoming (outgoing) photon polarization. The polar angle $\theta$ uniquely defines the direction of the scattered photon in the $xz$ plane (scattering plane). The target T is placed at the origin of the coordinate axes $xyz$.

%
\begin{figure}[t]
\begin{center}
\includegraphics[scale=0.35, trim=0cm 0cm 0cm 0cm, clip=true]{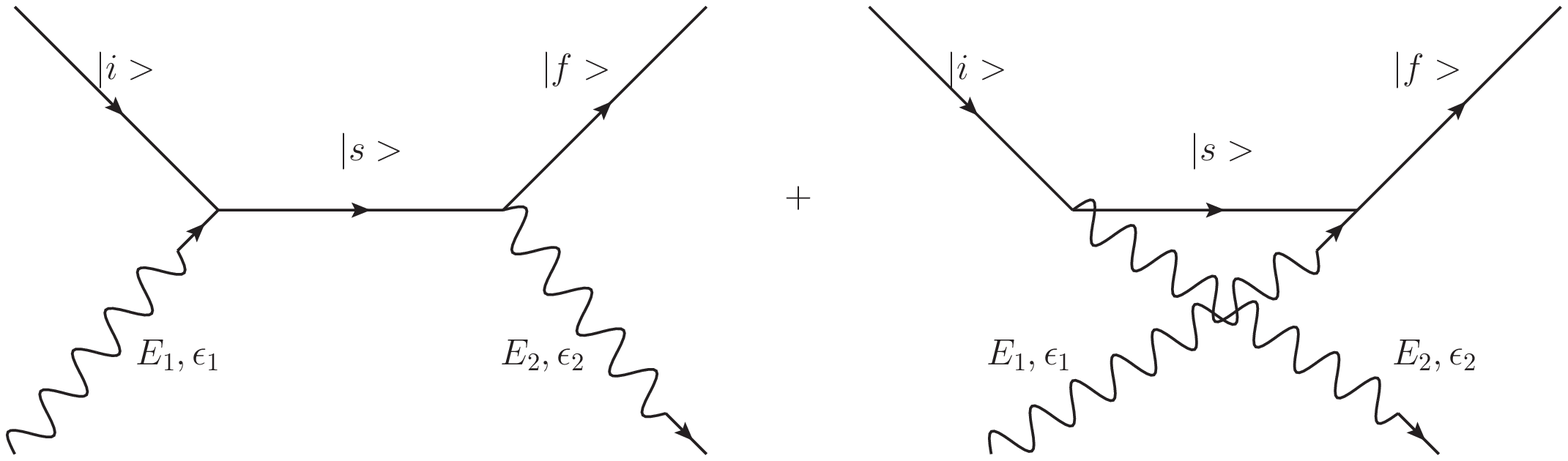}
\caption{Graphical representation of the scattering amplitudes $\mathcal{A}_{12}$ and $\mathcal{A}_{21}$. The state $\ket s$ is the target intermediate state during the scattering process.}
\label{fig:scattering}
\end{center}
\end{figure}

Each of the amplitudes $\mathcal{A}_{12, 21}$ describes one graph in Fig. \ref{fig:scattering}. 
The amplitude $\mathcal{A}\equiv\mathcal{A}_{12}+\mathcal{A}_{21}$ is the coherent sum of the two graphs. Both terms are in general challenging to compute, since both contain a summation over the infinite spectrum of the target bound system. 
Within the Thomson approximation, $\mathcal{A}_{12, 21}$ are approximated to zero, which is evidently justified from Eq. \eqref{eq:Aampl} if the photon energies ($E_{1,2}$) are much larger than the target binding energy.
The first term of the right member in Eq. \eqref{eq:Ampl} thus represents the Thomson scattering amplitude.
It must be therefore clear that any correction coming from the $\mathcal{A}$ term is to be regarded as a correction to the Thomson approximation. Note that the Thomson scattering amplitude is vanishing in the case of inelastic scattering (Raman scattering).

With the aim to avoid the summation over the (infinite) intermediate target states, we shall propose an alternative method that is based on a perturbation expansion.

\section{Perturbation expansion}
\subsection{Basic theory}
Let us focus on the term $\mathcal{A}_{12}$ and rewrite it as
\begin{equation}
\begin{array}{l}
\mathcal{A}_{12}\ds
=\ds\sum_{j,t=1}^N\bra{f}\opvector{p}_{j}\cdot \uvector{\epsilon}_2^*
\frac{1}{\opscalar H_0-E_i-E_1}\opvector{p}_t\cdot\uvector{\epsilon}_1\ket{i}~,
\end{array}
\label{eq:A1}
\end{equation}
where $\opscalar H_0$ is the target hamiltonian, and we used the eigenvalues equation $\opscalar H_0\ket{\nu}=E_\nu \ket{\nu}$ as well as the completeness of the states $\ket{\nu}$. Now, let us define the state $\ket{s}$ as
\begin{equation}
\ds \ket{s}\equiv\frac{c}{\opscalar{H}_0-E_i-E_1}\sum_{t=1}^N\opvector{p}_t\cdot\uvector{\epsilon}_1\ket{i}~,
\label{eq:Seq}
\end{equation}
where $c$ is the speed of light. Once the state $\ket{s}$ is known, the term $\mathcal{A}_{12}$ can be simply calculated as
\begin{equation}
\mathcal{A}_{12}=
\ds\frac{1}{c}\sum_{j=1}^N\bra{f}\opvector{p}_{j}\cdot \uvector{\epsilon}_2^*
\ket{s}~.
\label{eq:onephampl}
\end{equation}
The equation above resembles the amplitude for a single photon process, and can be calculated with high accuracy, provided that $\ket{s}$ is known. Our task is then to find a solution for the state $\ket{s}$. In the literature, there have been other studies that used a similar starting point to get inhomogeneous equations of the Green's function to be solved numerically \cite{dej1997, dad1992}. In contrast, we here seek an analytical solution of the problem via a perturbation expansion. 

From Eq. \eqref{eq:onephampl}, one sees that the state $\ket{s}$ evidently represents the quantum intermediate state of the target during the scattering process, as depicted in Fig. \ref{fig:scattering}.
With the aim to find such a state, we cast equation \eqref{eq:Seq} as
%
\begin{equation}
\ds\ket{s}=-\sum_{t=1}^N\frac{c\uvector{\epsilon}_1\cdot \opvector p_t}{E_1}\ket{i}+\frac{\opscalar H_0-E_i}{E_1}\ket{s}~.
\label{eq:EqS}
\end{equation}
Equation \eqref{eq:EqS} can be regarded as a perturbation expansion of the state $\ket{s}$ on $(\opscalar H_0-E_i)/E_1$. The term on which the expansion is made is easily interpretable: The numerator ($\opscalar H_0-E_i$) describes the energy shift caused by the scattering photon, while the denominator ($E_1$) is the incident photon energy. The expansion coefficient is thus a measure of how much energy shift is brought to the target by the scattering photon in units of the incident photon energy.

In the following, we shall explicitly calculate some perturbation orders. Note that the state $\ket{s}$ need not be normalized to one, since it does not represent a physical state. Close to resonances, from Eq. \eqref{eq:A1} it can be easily seen that $\braket{s}{s}\sim +\infty$.

\subsection{$0^{th}$ order}
\label{sec:ClassicCase}
At the $0^{th}$ order, it is assumed $1/E_1=0$, which entails $E_1\to+\infty$. From Eq. \eqref{eq:EqS}, this implies
\begin{equation}
\ket{s}_0=0 ~,
\end{equation}
where the subscript indicates the expansion order. In turn, this implies that the amplitudes $\mathcal{A}_{12,21}$ are identically zero at this expansion order, that is $\mathcal{A}_{12,21}^{(0)}=0$. The total scattering amplitude then turns out to be 
\begin{equation}
\mathcal{M}=\ds N \bra{f}\uvector{\epsilon}_1\cdot \uvector{\epsilon}_2^*\ket{i}= N~ \uvector{\epsilon}_1\cdot \uvector{\epsilon}_2^*~\delta_{i,f}~,
\label{eq:0thorder}
\end{equation}
where we used $\braket{f}{i}=\delta_{i,f}$. This expansion order evidently corresponds to the Thomson approximation.

One may notice from Eq. \eqref{eq:Ampl} that the same result can be accomplished by approximating $m\to +\infty$ and if the photon energy is far from the target spectrum. In other words, this is the case if the target can be considered classical. 
We may therefore conclude that approximating the intermediate state $\ket{s}$ at the $0^{th}$ order, i.e. $\ket{s}\approx\ket{s}_0$, is effectively like considering the target as classical. Because of this, we shall hereinafter call the $0^{th}$ order term in the perturbation expansion as the \supercomas{classical term}.

\subsection{$1^{st}$ order}
\label{sec:Analysis1st}
In order to compute the $1^{st}$ order of the perturbed state, we insert $\ket{s}_0$ into the right hand side of Eq. \eqref{eq:EqS}. We are thus approaching the exact scattering solution from the Thomson approximation, i.e. from photon energies above the spectrum. By doing so we get:
\begin{equation}
\begin{array}{lcl}
\ket{s}_1
&=&\ds-\sum_{t=1}^N\frac{c\uvector{\epsilon}_1\cdot \opvector p_t}{E_1}\ket{i}~.
\end{array}
\label{eq:s1st}
\end{equation}
It can be easily seen from Eq. \eqref{eq:EqS} that this approximation corresponds to considering $\opscalar H_0 \ket{s}\approx E_i \ket{s}$. Within this approximation, we are therefore considering as if the energy of the perturbed state $\ket{s}$ were approximately unperturbed, i.e. $E_s\approx E_i$.

By using \eqref{eq:s1st}, we can explicitly calculate the first order correction to the scattering amplitude:
$
\mathcal{A}_{12}^{(1)}=-\frac{1}{E_1} \sum_{j,t}\bra{f}
\opvector{p}_j\cdot \uvector{\epsilon}_2^*
\opvector{p}_t\cdot \uvector{\epsilon}_1
\ket{i}.
$
Analogously, the term $\mathcal{A}_{21}^{(1)}$ takes the form
$
\mathcal{A}_{21}^{(1)}=+\frac{1}{E_2} \sum_{j,t}\bra{f}
\opvector{p}_j\cdot \uvector{\epsilon}_1
\opvector{p}_t\cdot \uvector{\epsilon}_2^*
\ket{i}.
$ 
Without restriction of generality and for simplicity, let us consider the case for which photon polarizations are measured in the linear basis, for which $\uvector\epsilon_{1(2)}^*=\uvector\epsilon_{1(2)}$. By using $\comm{\opvector{p}_j\cdot \uvector{\epsilon}_2}{\opvector{p}_t\cdot \uvector{\epsilon}_1}=0$ for any $j$, $t$, we obtain
\begin{equation}
\begin{array}{l}
\mathcal{A}^{(1)}=\mathcal{A}_{12}^{(1)}+\mathcal{A}_{21}^{(1)}\\[0.4cm]
\ds\quad = \sum_{j,t=1}^N\bra{f}
\opvector{p}_j\cdot \uvector{\epsilon}_1 \opvector{p}_t\cdot \uvector{\epsilon}_2
\ket{i}
\left(
\frac{1}{E_2}
-
\frac{1}{E_1}
\right)~.
\end{array}
\label{eq:RamanAmpl}
\end{equation}
This is a fundamental correction to the Thomson amplitude that depends only on initial and final states. The scattering operator is in fact independent of the target binding potential. Given that the Thomson amplitude is vanishing for Raman scattering, equation \eqref{eq:RamanAmpl} is actually the first non-vanishing quantum mechanical term related to Raman processes.

A first remark from Eq. \eqref{eq:RamanAmpl} is that such equation could be also directly obtained from Eq. \eqref{eq:Aampl} by considering $E_{1,2}\gg E_{i,f,\nu}$ and by then using the completeness of the intermediate states, $\sum_\nu \ket\nu\bra\nu =1$. A second remark is that the amplitude goes to zero as $E_1\to E_2$. That demonstrates that the first order correction is always zero in Rayleigh scattering, independently of the binding potential. This rule is a consequence of the coherence between incoming and outgoing light in Rayleigh scattering. 

\subsection{$n^{th}$ order}
To find the $2^{nd}$ order of the perturbed state, we insert the $1^{st}$ order solution into the right hand side of Eq. \eqref{eq:EqS}:
\begin{equation}
\begin{array}{lcl}
\ket{s}_2&=&\ds\ket{s}_1-\frac{1}{E_1^2}\left[\opscalar H_0,\sum_{t=1}^N c\uvector{\epsilon}_1\cdot \opvector p_t\right]\ket{i}~,
\end{array}
\label{eq:2ndorder}
\end{equation}
where we used $\big(\opscalar H_0-E_i\big)\ket{i}=0$ and the fact that $E_i$ commutes with any quantum operator. 
By $n$ replacements, the state at order $n$ is found to be
\begin{equation}
\ds\ket{s}_{n}=\ds\left\{
\begin{array}{lr}
0&\textrm{ for $n=0$}\\
\ds \ket{s}_{n-1}+\frac{\opscalar O_n}{(E_1)^n}\ket{i}&\textrm{ for $n>0$}
\end{array}
\right.
\label{eq:exp}
\end{equation}
or equivalently
\begin{equation}
\ds\ket{s}_{n}=\left\{
\begin{array}{lr}
0&\textrm{ for $n=0$}\\
\ds\sum_{k=1}^n\frac{\opscalar O_k}{(E_1)^k}\ket{i}
&\textrm{ for $n>0$}
\end{array}
\right.
\label{eq:exp2}
\end{equation}
where 
\begin{equation}
\begin{array}{lcl}
\opscalar O_{k} &=& -\frac{im}{\hbar}\sum_t c\uvector\epsilon_1\cdot\overbrace{\Big[
\opscalar H_0, \big[
\opscalar H_0,[...,[\opscalar H_0}^{\substack{\opscalar H_0\textrm{ repeated for }\\k\textrm{ times}}},\opvector r_{t} ]
...]\big]
\Big]\\[0.4cm]
&=&\ds \comm{\opscalar H_0}{\opscalar O_{ k-1}}
~,
\end{array}
\label{eq:commRule}
\end{equation}
while $\opscalar O_{0}=-\frac{im}{\hbar}\sum_t c\uvector\epsilon_1\cdot\opvector r_{t}$ and $\opscalar{O}_1=-c\uvector\epsilon_1\cdot\sum_t\opvector p_t$. 
To write Eq. \eqref{eq:commRule}, we used the equivalence $\opvector p_t=\frac{im}{\hbar}\comm{\opscalar H_0}{\opvector r_t}$, which holds as long as the target binding potential commutes with the position operator.
From Eq. \eqref{eq:commRule}, one sees that, if $\opscalar O_{k-1}$ is vanishing, then also $\opscalar O_{k}$ is vanishing, as well as all operators $\opscalar O_{j}$ with $j\ge k$.

We define $\opscalar{T}$ as the {\it transition operator}, that is the operator that transforms the initial state into the intermediate state, viz. 
\begin{equation}
\ket{s}\equiv\opscalar{T}\ket{i}~.
\end{equation}
From Eqs. \eqref{eq:exp}, \eqref{eq:exp2}, it immediately follows
\begin{equation}
\opscalar{T}=\frac{\opscalar{O}_1}{E_1}+
\frac{\opscalar{O}_2}{E_1^2}+\frac{\opscalar{O}_3}{E_1^3}+...~,
\label{eq:Texp}
\end{equation}
Therefore, the transition operator at order $n$ is given by $\opscalar{T}_n=\sum_{k=1}^n\frac{\opscalar O_k}{(E_1)^k}$.
Equations \eqref{eq:exp}-\eqref{eq:Texp} can be used jointly with Eqs. \eqref{eq:onephampl} and \eqref{eq:Ampl} to find the total two-photon scattering amplitude, given the target hamiltonian $\opscalar{H}_0$. We shall explicity do this in Secs. \ref{sec:SimpleCases} and \ref{sec:MultiCases}.


\section{General theoretical results and remarks}
By using the theory presented in the previous sections, we here derive several general results in the form of statements and formulas, which are potentially useful when analyzing Rayleigh and Raman scattering. We shall also use them later in Secs. \ref{sec:SimpleCases} and \ref{sec:MultiCases}.

\subsection{Scattering cross section formula}
The dependence of the scattering amplitude on the photon energies $E_{1,2}$ is wholly within the denominators of the kind $\sim 1/E_{1,2}$ that are contained in the transition operator $\opscalar T$ (see Eq. \eqref{eq:Texp}). This means that, without any assumption on the target binding potential, we may predict the cross section dependence on the photon energy. For example, let us take Raman scattering, for which $\braket{f}{i}=0$ and $E_i\neq E_f$. While the energy of the incident photon $E_1\equiv E$ is freely adjustable, the energy of the scattered photon is bound by energy conservation to be $E_2=E-E_{res}$, where we defined the resonance energy $E_{res}=E_f-E_i$. The cross section at the leading order of the series expansion is thus proportional to (see Eq. \eqref{eq:RamanAmpl})
\begin{equation}
\sigma^{(1)}\propto \left|\mathcal{M}^{(1)}\right|^2\propto \left|\frac{1}{E_2} - \frac{1}{E_1}\right|^2=
\frac{E_{res}^2}{E^2}
\frac{1}{\big(E-E_{res}\big)^2}~.
\label{eq:sigma1}
\end{equation}
Deviations from this formula come from higher orders in $1/E_{1,2}$. For example, from Eq. \eqref{eq:Texp} the cross section at the second order can be written as
\begin{equation}
\begin{array}{l}
\sigma^{(2)}\propto  \ds\left|\mathcal{M}^{(2)}\right|^2\propto 
\left|
c_1\left(\frac{1}{E_2} - \frac{1}{E_1}\right)
+
c_2\left(\frac{1}{E_2^2} + \frac{1}{E_1^2}  \right)
\right|^2\\[0.4cm]
\approx \ds 
\left(\frac{E_{res}^2}{E^2}
\frac{|c_1|^2}{\big(
E-E_{res}
\big)^2}
+
 \frac{E_{res}}{E}\frac{4\Re(c_1 ~ c_2^* )}{(E-E_{res})^3}
\right)~,
\end{array}
\label{eq:sigma2}
\end{equation}
where we assumed $|c_2|\ll |c_1|$, and we kept terms of lowest order in $1/E$. 
The coefficients $c_{1,2}$ are scaling factors that depend on the matrix elements of the transition operator (and therefore they depend in general on the target potential).
The second term in the right hand side of Eq. \eqref{eq:sigma2} represents the first correction to the leading order term. One may therefore measure the energy dependence of the scattering cross section and parametrize it as in Eq. \eqref{eq:sigma2} (and subsequent orders), by using the coefficients $c_{1,2,3,...}$. This will help isolating the quantum contributions to the scattering cross section, and would also provide an empirical cross section formula whose terms have physical meaning. From the fitted coefficients, one would then be able to retrieve the (dipole) matrix elements for the target specimen. 

\subsection{Cancellation of quantum contributions for targets composed by identical particles}
\label{sec:Cancellation}

Let us suppose the target to be composed by just two particles of equal mass and charge, which we denote by particle $A$ and $B$. Suppose also that such two particles experience a two-body potential that depends on the reciprocal distance, that is of the type $V(\bm r_A - \bm r_B)$, as it is mostly the case in nature, e.g. the Coulomb potential. 
Calculating the second order contribution of the operator $\opscalar O$ we obtain
\begin{align}
\opscalar{O}_2&= \ds\comm{\sum_{j=A,B}\frac{\opvector p^2_j}{2m} + V(\opvector r_A-\opvector r_B)}{\opscalar O_1}
\nonumber\\[0.5cm]&
	=-c \uvector \epsilon_1 \cdot \ds\sum_{j=A,B}\Big[V(\opvector r_A-\opvector r_B),\opvector p_j\Big]\nonumber\\[0.5cm]
	&=-i\hbar c\uvector \epsilon_1 \cdot \ds\sum_{j=A,B}\bm\nabla_{\bm r_j} ~ V(\opvector r_A-\opvector r_B)
	~.
\label{eq:secondOff}
\end{align}
However, since $\bm\nabla_{\bm r_B} V(\opvector r_A-\opvector r_B)=-\bm\nabla_{\bm r_A} V(\opvector r_A-\opvector r_B)$, then $\opscalar{O}_2=0$. Consequently, $\opscalar{O}_{k\ge 2}=0$, as seen from Eq. \eqref{eq:commRule}. Therefore the only two non-vanishing terms of the perturbation expansion are the classical term \eqref{eq:0thorder} and the first quantum correction \eqref{eq:RamanAmpl}. If light is scattered elastically (as it is mostly the case), then also the quantum correction is vanishing. Therefore the target behaves as a classical scatterer, since all quantum contributions are identically zero. This result can be trivially extended to a target composed by any number of identical particles, as long as the inter-particle potentials are reciprocal. 

Summarizing, this finding shows that when the target is composed by identical particles that interact with light, the quantum contributions to the scattering amplitude are either zero or significantly reduced. More specifically, the dependence of the inter-particle potential on the reciprocal distance generates coherent scattered waves that interacts destructively with each other, in pair, thus resulting in a cancellation of the quantum contribution to the scattering amplitude. In case of elastic scattering (Rayleigh scattering), the cancellation of quantum contributions is complete. The resulting scattered wave is thus the same as if it were scattered by a classical target. Hence, 
the process of elastic scattering does not retrieve any information about the quantum nature of the scatterer.
On the other hand, in the case of inelastic scattering (Raman scattering) the cancellation of quantum terms is almost complete, since only one term is left out of the perturbation expansion, beside the classical term.

One could use this result in different areas, ranging from fundamental to applied physics. For example, one could build quantum information carriers made of identical bound particles that interact with light, such as BCS pairs \cite{devoret2013superconducting}. The quanta of information could be embedded into a quantum feature of the bound system that is not retrievable by the classical term. Thus, any attempt to steal the information from the system with elastic light scattering would fail, provided that dipole approximation is valid. Moreover, this result has also impact on the coherence time of the quantum carriers, which is an ongoing research field \cite{paik2011observation, barends2013coherent}, since it predicts a suppression of electromagnetic noise for carriers made of identical particles.

Potentials that depend on the reciprocal distance are typical in atoms and nuclei. In atoms, however, electrons experience also interaction with the nucleus, other than with themselves. Similarly, in nuclei protons experience interactions with neutrons, other than with themselves. The resulting overall potential is thus not only among particles of equal mass and charge, but also among particles with different mass and charge. Because of this, the quantum cancellation does not fully apply, and consequently the scattered light does possess information about the quantum nature of the target. 
Nevertheless, we can show that there is anyways a suppression of the contribution of reciprocal potentials among identical particles, such as electron-electron or proton-proton repulsion, at high energies. To this aim, let us call $\opscalar V_R\equiv\sum_{\mu> \nu}V\left(\opvector r_{\mu}-\opvector r_{\nu}\right)$ the reciprocal potential of the identical compounds within the target, where $\mu,\nu=(1, ..., N)$ indexes the compound. One may straightforwardly compute the scattering operators $\opscalar O_1$, $\opscalar O_2$, $\opscalar O_3$, and find out that they are linear in the momentum operator. They therefore commute with the reciprocal potential, as showed above. The fourth order is the lowest order where the scattering operator presents non-linear terms in the momentum operator. As a matter of fact, $\opscalar O_4$ presents terms of the type $\sim \opscalar p_{\mu i}~\opscalar p_{\nu j}$, where $i,j=(x,y,z)$ are the cartesian coordinates. At the fifth order, the scattering operator gets non-zero contributions from the reciprocal potential since in general $\opscalar O_5\propto \comm{\opscalar V_R}{\opscalar O_4}\neq 0$. Therefore, the fifth order is the lowest scattering order in which the reciprocal potential within identical particles in the target, such as electron-electron or proton-proton repulsion, contributes to the scattering cross section. This entails that the contribution of reciprocal potentials to the scattering cross section is reduced in those cases, as long as the photon energy is high enough to lead to convergency in Eq. \eqref{eq:exp}. In addition to this argument, we shall demonstrate in Sec. \ref{sec:MultiCases} that reciprocal potentials do not contribute at all to the Rayleigh scattering amplitude, as long as a) the binding potential can be approximated to harmonic, and b) the target wavefunction can be separated into relative and center-of-mass coordinates.

\subsection{Information retrieved by linearly polarized light}
\label{sec:LinearInfo}
Let us suppose that the target binding potential is of the form $V(x,y,z)=V(x)+V(y)+V(z)$, and that the incident light is linearly polarized along the $x$-direction. From Eq. \eqref{eq:commRule}, we can easily calculate that the perturbation equation for the scattering operator $\opscalar O$ depends only upon the potential along the $x$-axis:
\begin{align}
\opscalar O_k &=\ds \Big[
\frac{\opvector p^2}{2m}+V(\opscalar x,\opscalar y,\opscalar z),\Big[...,\Big[\frac{\opvector p^2}{2m}+V(\opscalar x,\opscalar y,\opscalar z), -c\opscalar p_x\Big]
...\Big]\Big]\nonumber\\[0.6cm]
&=\ds \Big[
\frac{\opscalar p_x^2}{2m}+V(\opscalar x),\Big[...,\Big[\frac{\opscalar p^2_x}{2m}+V(\opscalar x), -c\opscalar p_x\Big]
...\Big]\Big]
~.
\end{align}
This entails that the total cross section will only depend on the potential along the $x$-axis. 
Generalizing, if the target binding potentials along different cartesian axes are fully decoupled (i.e. they are different terms in the hamiltonian), linearly polarized light will {\it only} probe the binding potential along the polarization axis. 

This result is particularly useful if one wants to individually probe the target binding potential along a given axis, as it might be the case, for example, in ion traps \cite{leibfried2003quantum}, especially in array-based approaches where the transport of ions along the trap axis is sought, to be eventually used for quantum communication purposes \cite{kielpinski2002architecture}. 
Alternatively, this result can be used to polarize light, by inducing asymmetric target potentials, where the asymmetry needs to be in the polarization plane. 
Conversely, one could retrieve asymmetries in the target binding potential by analyzing the polarization of the scattered light.

\subsection{Further considerations}
Thanks to the commutators, at the $n^{th}$ order we get a correction proportional to $\hbar^n$. This feature resembles typical semiclassical expansions which are also in powers of $\hbar$, such as the WKB approximation \cite{sakurai2011}. Here, however, the inverse dependence on the photon energy contributes to the convergence of the series in addition to the dependence on $\hbar$. If the target binding potential and all its derivatives are not singular anywhere, the convergence of the series in Eqs. \eqref{eq:exp}-\eqref{eq:Texp} will be attained, at least for energies significantly above the target spectrum. To this regard, one must be careful when considering Coulomb binding potentials since they present singularities given by terms $\propto 1/|\bm r_i|$ or $\propto 1/|\bm r_i-\bm r_j|$. 

As last remark, we point out that an exact solution of the series expansion in Eq. \eqref{eq:commRule} is in general difficult to find. Even so, below we shall show a few cases for which a closed solution can be found. Moreover, we shall show that convergence for any (arbitrarily complex, not singular) binding potential is guaranteed provided that the photon energy is high enough with respect to the binding energy. This endows the formulas \eqref{eq:exp}-\eqref{eq:commRule} with a fundamental value that paves the way to compute two-photon processes within a full second-order quantum mechanical framework for complex quantum systems where the infinite sum of intermediate states is difficult or unfeasible to calculate.

\section{Application to a few single-particle cases}
\label{sec:SimpleCases}
As practical examples, in this section we shall apply the technique we developed above to some specific (single-particle) cases that are easily found in literature for modeling quantum phenomena. We shall proceed in order of complexity with regard to the target binding potential. We will analyze Raman or Rayleigh light scattering, depending on the case study.

\FloatBarrier
\subsection{Potential box}
\label{sec:PotBox}

The simplest binding potential is a potential box, where the potential is zero (or any constant) within the box, while it is infinite outside. Notwithstanding its simplicity, such a potential is used to model bound states in nuclear physics \cite{wilets1958theories}, subnuclear physics \cite{chodos1974new}, and semi-conductor physics \cite{kolbas1984man}, among others. Eigenstates and eigenenergies can be found in standard textbooks \cite{brajo1983}. Specifically, the combination of quantum numbers $(n_x, n_y, n_z)$ defines the eigenstates, with $n_j=1,2,3,...$ for any $j=x,y,z$. Quantum states are characterized by the energy $E_{n}=\xi_a n^2$, where $n^2=n_x^2+n_y^2+n_z^2$ and $\xi_a=\frac{\pi^2\hbar^2}{2 m a^2}$. 
Given that all eigenfunctions are vanishing at the edge of the potential box and beyond, we can restrict the integrals in Eq. \eqref{eq:Aampl} within the domain $|x|<a/2$, $|y|<a/2$, $|z|<a/2$, where $a$ is the size of the potential box. Within such domain, the potential is constant and therefore $\opscalar{O}_{k\ge 2}=0$. By using the theory developed in the previous sections, the total (exact) scattering amplitude $\mathcal{M}$ is calculated as
\begin{equation}
\begin{array}{lcl}
\ds\mathcal{M}&=&\ds\uvector{\epsilon}_1\cdot \uvector{\epsilon}_2~\delta_{i,f}
-\frac{1}{m}
\left(\frac{1}{E_2}-\frac{1}{E_1}\right)
\underbrace{\bra{f}\opvector{p}\cdot \uvector{\epsilon}_1 \opvector{p}\cdot \uvector{\epsilon}_2\ket{i}}_{C_{f,i}}
\end{array}
\label{eq:PotBoxM}
\end{equation}
where $E_2=E_1-E_f+E_i$ by energy conservation, and where we considered (without restriction of generality) linear photon polarizations. The matrix elements $C_{f,i}$ can be calculated by using standard techniques. 

\begin{figure}[t]
\includegraphics[scale=0.85]{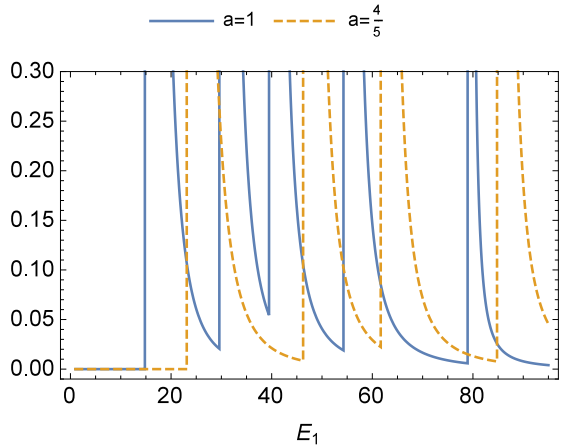}
\caption{
The probability density, $|\mathcal{M}|^2$, for Raman scattering off one particle trapped by a potential box (see Sec. \ref{sec:PotBox}).
Atomic units are used.
The potential parameters can be retrieved by analyzing the relative positions of the resonance peaks. 
}
\label{fig:BoxPotential}
\end{figure}
Let us investigate Raman scattering off one particle trapped in the potential box.
Provided that the matrix element $C_{f,i}$ is not zero, the Raman cross section will peak at photon energy $E_1\simeq E_f-E_i$, as can be seen from Eq. \eqref{eq:PotBoxM}. Since $\xi_a$ depends on $a$, the relative position of the energy peaks will also depend on $a$ (the box size). For example, the first peak is located at energy $E_1=3\xi_a=3\pi^2\hbar^2/(2ma^2)\approx 15/a^2$ (atomic units). By scanning through the incident photon energy and by thus investigating the distance between peaks, one can retrieve the size of the potential box. See Fig. \ref{fig:BoxPotential} for this purpose, where $\abs{\mathcal{M}}^2$ is plotted after having summed over the final target states, integrated over the scattering angle, as well as summed (averaged) over the final (initial) photon polarizations.
Furthermore, the dependence of the energy peaks on the box size $a$ can be used to roughly estimate the range of the target binding potential, as long as it can be approximated to a potential box.

A similar analysis can be performed for a semi-infinite potential well or for a delta potential, which are potentials that can be used to model nucleon-nucleon and short-range interactions, respectively \cite{abdullah2017fundamentals}.

\FloatBarrier
\subsection{Symmetric linear potential}
\label{sec:Linear}
\begin{figure}[t]
\includegraphics[scale=0.85]{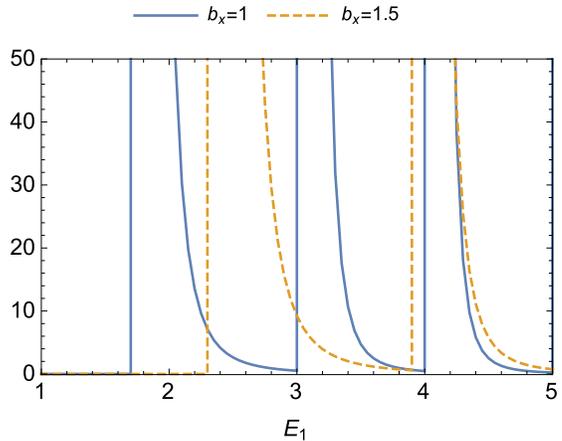}
\caption{
Same as Fig. \ref{fig:BoxPotential}, but relative to the case-study presented in Sec. \ref{sec:Linear}.
}
\label{fig:LinearPotential}
\end{figure}

Besides box potentials, linear potentials are also a class of potential models that are used in different areas of physics \cite{pauli2003hadronic, colella1975observation, han2015dielectric, fratini2014quantum}. Let us therefore consider light scattering off one-particle bound by a potential of the form $V=b_x|x|+b_y|y|+b_z|z|$, where $b_{x,y,z}$ are constants. Eigenstates and eigenenergies are well known and can be found in literature \cite{Glasser2015}. By using linear photon polarizations (without restriction of generality), from Eq. \eqref{eq:commRule} we can solve for the operator $\opscalar O$ of the amplitude $\mathcal{A}_{12}$:
\begin{equation}
\begin{array}{lcl}
\opscalar O_2&=&\ds- i\hbar c \uvector \epsilon_{1} \cdot \bbarOp~,
\hspace{0.5cm} \opscalar O_{k\ge 3}=0 ~,
\end{array}
\label{eq:BoxEqs}
\end{equation}
where $\bbarOp=(	\opscalar x b_x/|\opscalar x|, \opscalar y b_y/|\opscalar y|, \opscalar z b_z/|\opscalar z|)$,
while $\opscalar O_1$ has been defined in Eq. \eqref{eq:commRule}. Equations \eqref{eq:BoxEqs} are valid only for $x\neq 0$, $y\neq 0$, $z\neq 0$. On the other hand, if $x=0$ or $y=0$ or $z=0$, the operators $\opscalar O_{k\ge 2}$ cannot be calculated since the first derivative of the potential is not defined. 
Nevertheless, this problem can be circumvented by either taking the principal value of the integral in 
Eq. \eqref{eq:onephampl}, or by working with antisymmetric target wavefunctions, which are vanishing in those points.

By coherently summing the amplitudes $A_{12}$ and $A_{21}$ as in Eq. \eqref{eq:Ampl}, the total (exact) scattering amplitude $\mathcal{M}$ is found as
\begin{equation}
\begin{array}{lcl}
\ds\mathcal{M}&=&\ds\uvector{\epsilon}_1\cdot \uvector{\epsilon}_2~\delta_{i,f}
-\frac{1}{m}
\left(\frac{1}{E_2}-\frac{1}{E_1}\right)
\bra{f}\opvector{p}\cdot \uvector{\epsilon}_1 \opvector{p}\cdot \uvector{\epsilon}_2\ket{i}\\[0.4cm]
&&\ds -\frac{i\hbar}{m}\bra{f}\left(
	\frac{\uvector \epsilon_2 \cdot \opvector p ~ \uvector \epsilon_1 \cdot \bbarOp}{E_1^2}
	+
	\frac{\uvector \epsilon_1 \cdot \opvector p ~ \uvector \epsilon_2 \cdot \bbarOp}{E_2^2}
\right)\ket{i}~,
\end{array}
\label{eq:SimmLinM}
\end{equation}
where $E_2=E_1-E_f+E_i$ by energy conservation. 

Let us consider Raman scattering, where the initial state is the (symmetric) ground state, while the incident photon linear polarization is along the $x$-axis ($\chi_1=0$). Let us further consider for simplicity the case of a) scattering in the backward direction ($\theta=\pi$), and b) no linear polarization flip, which implies $\chi_2=0$. Figure \ref{fig:LinearPotential} displays $\abs{\mathcal{M}}^2$, after having summed over the final target states. As for the previous case, the relative distance between peaks can be used to retrieve the potential parameter $b_x$. On the other hand, by virtue of the chosen settings, the scattering amplitude does not depend on the other two parameters $b_y$, $b_z$, which is also a consequence of our findings in Sec. \ref{sec:LinearInfo}.

\FloatBarrier
\subsection{Harmonic potential}
\label{sec:sampleH}
Here we consider the target being characterized by one charged particle bound by a harmonic potential. Such a model is extensively used in quantum optics \cite{you2011atomic, moras1996semiconductor} as well as in nuclear physics (shell model) \cite{caurier2005shell}, among others.
The target hamiltonian is 
\begin{equation}
\opscalar{H}_{0}=\opscalar{H}_{ho}=\frac{\opvector{p}^2}{2m}+\frac{m\omega^2}{2}(\opvector{r}-\bm{r_0})^2~,
\end{equation}
where $\omega$ is the oscillator constant, and $\bm{r_0}$ the displacement vector. Let us further consider an incident photon that is linearly polarized along the $x$-direction, which entails $\uvector{\epsilon}_1=(1,0,0)$. Consequently, we have $\opscalar O_1=-c\opscalar{p}_x$. By straightforward calculation, one can see that$
\comm{\opscalar{H}_{0}}{\opscalar O_1}=-c\omega^2 (\opscalar{x}-x_0) i\hbar m
$, as well as $
\comm{\opscalar{H}_{0}}{\comm{\opscalar{H}_{0}}{\opscalar O_1}}=\hbar^2\omega^2\opscalar O_1
$. This can be replaced in the definition of $\opscalar{T}$ to find
\begin{widetext}
\begin{align}
\opscalar{T}&=
\frac{\opscalar O_1}{E_1}+
\frac{\comm{\opscalar{H}_{0}}{\opscalar O_1}}{E_1^2}+
\frac{\comm{\opscalar{H}_{0}}{\comm{\opscalar{H}_{0}}{\opscalar O_1}}}{E_1^3}+
\frac{\comm{\opscalar{H}_{0}}{\comm{\opscalar{H}_{0}}{\comm{\opscalar{H}_{0}}{\opscalar O_1}}}}{E_1^4}+
...\nonumber\\
&=\frac{\opscalar O_1}{E_1}+\frac{\comm{\opscalar{H}_{0}}{\opscalar O_1}}{E_1^2}+
\frac{\hbar^2\omega^2}{E_1^2}\left(
\frac{\opscalar O_1}{E_1}+
\frac{\comm{\opscalar{H}_{0}}{\opscalar O_1}}{E_1^2}+
\frac{\comm{\opscalar{H}_{0}}{\comm{\opscalar{H}_{0}}{\opscalar O_1}}}{E_1^3}+
...
\right)\nonumber\\
&=\frac{\opscalar O_1}{E_1}+\frac{\comm{\opscalar{H}_{0}}{\opscalar O_1}}{E_1^2}
+\frac{\hbar^2\omega^2}{E_1^2}\opscalar{T}~.
\label{eq:TinHO}
\end{align}
\end{widetext}
This leads to a closed solution:
\begin{align}
\left(1- \frac{\hbar^2\omega^2}{E_1^2}  \right)\opscalar{T}=\frac{\opscalar O_1}{E_1}+\frac{\comm{\opscalar{H}_{0}}{\opscalar O_1}}{E_1^2}~.
\label{eq:TfinalUND}
\end{align}
In the case of zero displacement ($|\bm{r_0}|=0$), the equation above can be recast as
\begin{align}
\opscalar{T}&=-i\frac{\hbar^2\omega^2}{E_1^2-\hbar^2\omega^2}\sqrt{\frac{2mc^2}{\hbar\omega}}\opscalar{a}^\dagger_{x, \gamma}~,
\label{eq:Tfinal}
\end{align}
where $\opscalar{a}^\dagger_{x, \gamma}=\sqrt{\frac{m\omega}{2\hbar}}\left(
\opscalar x-\frac{i}{m\omega}\gamma\opscalar{p}_x
\right)$
=
$\frac{1}{2}\big(\opscalar{a}^\dagger_x(1+\gamma)+\opscalar{a}_x(1-\gamma)\big)$ 
and $\gamma=E_1/(\hbar\omega)$. Here, $\opscalar{a}_x$ and $\opscalar{a}^\dagger_x$ represent the standard annihilation and creation operator for the quantum harmonic oscillator along the $x$-direction \cite{sakurai2011}.
While $\gamma$ denotes the ratio between the photon energy and the oscillator energy, $\opscalar{a}^\dagger_{x, \gamma}$ can be considered the `perturbed' creation operator along the polarization direction. 
This leads us to an additional result: when linearly polarized photons with energy $E_1$ are scattered by a harmonic oscillator with angular frequency $\omega$, the intermediate scattering state of the harmonic oscillator is equal to $\ket{s}=\opscalar{T} \ket{i}=K\opscalar{a}^\dagger_{j, \gamma}\ket{i}$, where $j$ is the photon polarization direction and $K$ is a numerical factor defined from Eq. \eqref{eq:Tfinal}. We may notice here again that, if the photon hits the resonance ($E_1\to \hbar\omega$), then $K\to +\infty$, and therefore $\braket{s}{s}\sim +\infty$.

From Eqs. \eqref{eq:Tfinal} and \eqref{eq:onephampl}, the term $\mathcal{A}_{12}$ can be calculated. Considering a general linear polarization for the incident photon, one has
\begin{align}
\label{eq:A12HO}
\mathcal{A}_{12}&=
m\left(\frac{\hbar^2\omega^2}{E_1^2-\hbar^2\omega^2}\right)
\sum_{\substack{j,k=\\x,y,z}}
\epsilon_{2j}\epsilon_{1k}
\bra{f}\big(
\opscalar{a}^\dagger_j-\opscalar{a}_j
\big)\opscalar{a}^\dagger_{k\gamma}
\ket{i}~.
\end{align}

\begin{figure}[t]
\includegraphics[scale=0.85]{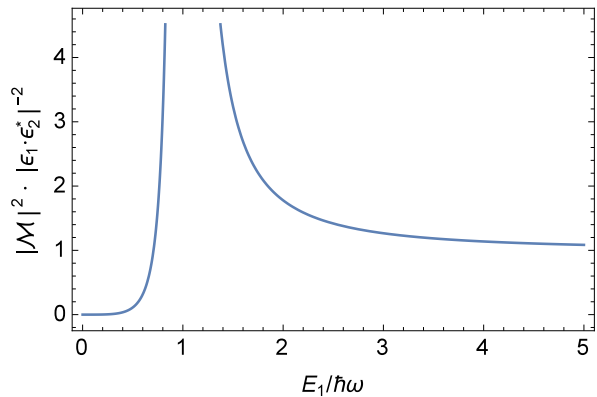}
\caption{
Probability density for elastic light scattering (Rayleigh scattering) off a particle bound by a harmonic potential.  
}
\label{fig:HORay}
\end{figure}

Finally, with the help of Eq. \eqref{eq:Ampl}, in the case of Rayleigh scattering (i.e. for $E_1=E_2\equiv E$, which implies $\ket{i}=\ket{f}$), the total amplitude $\mathcal{M}$ (exact) can be evaluated analytically:
\begin{equation}
\mathcal{M}=\uvector{\epsilon}_1\cdot\uvector{\epsilon}_2^*~\left(
1+\frac{\hbar^2\omega^2}{E^2-\hbar^2\omega^2}
\right)~.
\label{eq:FinalMHO}
\end{equation}
It can be easily noticed that the second term (which is the quantum term) represents a Lorentzian peak, with zero resonance width, the probability density being proportional to $\frac{\hbar^4 \omega^4}{(E+\hbar\omega)^2}\frac{1}{(E-\hbar\omega)^2}$.
The fact that the resonance width is zero is not unexpected, since we have not inserted the widths of bound states into the formalism. Moreover, we may notice that the amplitude peaks at the resonance $E=\hbar\omega$, which is the energy gap between neighboring states in a harmonic potential spectrum. This is easily understandable, since within the dipole approximation the transition operator is proportional to $\uvector\epsilon\cdot\opvector{p}$, and thus proportional to $\uvector\epsilon\cdot(\opvector{a}^\dagger-\opvector{a})$. Therefore, such operator has non-vanishing matrix elements only between states whose energy difference is $\hbar\omega$. One could also use this feature to retrieve Eq. \eqref{eq:FinalMHO} directly from Eqs. \eqref{eq:Aampl} and \eqref{eq:Ampl}, by restricting the summation over intermediate states to neighboring states.
In Fig. \ref{fig:HORay} we show the probability density for Rayleigh scattering off a harmonic oscillator with angular frequency $\omega$, as obtained from \eqref{eq:FinalMHO}.

\medskip

We see from Eq. \eqref{eq:FinalMHO} and from Fig. \ref{fig:HORay} that the amplitude $\mathcal{M}$ asymptotically vanishes as $E \to 0$. At low photon energies, the contribution of higher terms (the quantum terms, which are represented by the second addend in Eq. \eqref{eq:FinalMHO}), fully destructively interferes with the contribution of the zero order term (the classical term, which is represented by the first addend in Eq. \eqref{eq:FinalMHO}). Overall, this results in a vanishing cross section. At low light frequency, also Raman scattering is zero, since the light does not carry enough energy to excite the target. Thus, a harmonic oscillator does not scatter light when the light frequency is much lower than the oscillator frequency. 
This is in line with the Rayleigh scattering formula, which states that the scattering cross section for low energetic light is proportional to the fourth power of the light frequency \cite{hechtoptics}.

\subsection{Morse potential}
\label{sec:Morse}

The Morse potential is typically used in molecular physics to model vibrations \cite{Dong2007, Kader2002}. In this section we consider the target having one charged particle that is vibrating along the $x$-axis, the vibration being modeled by a Morse potential 
\begin{equation}
V(x)=V_0(e^{-2a(x-x_0)}-2e^{-a(x-x_0)})~,
\end{equation}
where $x$ and $x_0$ are respectively the position with respect to the core potential and the position at the equilibrium, while $a$, $V_0$ are parameters. The eigenstates of this potential are known \cite{Dong2007}. Let us denote by $\mathcal{V}(y, z)$ the binding potential along directions different than $x$. Let us also suppose that the incident light is linearly polarized along the $x$-direction. In this situation, $\mathcal{V}(y, z)$ does not contribute to the scattering amplitude, as showed in Sec. \ref{sec:LinearInfo}. We are thus selectively probing the Morse potential axis with light \cite{Dixneuf2015}.

\begin{figure}[t]
    \includegraphics[scale=0.55]{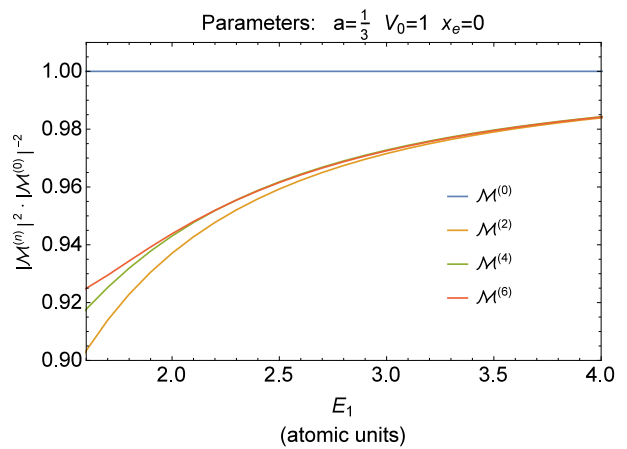}
    \caption{
	Probability density of elastic light scattering (Rayleigh scattering) off a particle vibrating along the $x$-direction, the vibration being modeled by a Morse potential. Light is polarized along the $x$-direction. Parameters are set as showed. $\mathcal{M}^{(n)}$ means scattering amplitude evaluated at the $n^{th}$ order, with $\mathcal{M}^{(0)}=\uvector  \epsilon_1\cdot \uvector \epsilon_2^*$. Convergence of the series expansion can be noticed.
	}
	\label{fig:MorseRay}
\end{figure}

We shall consider Rayleigh scattering off the ground state. The calculated components of the scattering operator $\opscalar O$, at orders 0 to 6, are showed in Appendix. Convergence of the series \eqref{eq:exp} is attained at sufficiently high energies. We explicitly show such a convergence in Fig. \ref{fig:MorseRay}, where the scattering probability density is plotted for different perturbation orders, for the specific case of a) scattering in the backward direction ($\theta=\pi$), and b) no linear polarization flip, which implies $\chi_2=0$. 

The elastic resonance peak does not appear in Fig. \ref{fig:MorseRay}, since it is located at lower energies. In the energy range related to the resonance peak, convergence of the perturbation expansion is difficult to attain, since it requires computing many expansion orders. Generally, convergence of the perturbation expansion is attained within few expansion orders if the photon energy is sufficiently above the resonance energies. 
%

\FloatBarrier
\section{Application to multi-particle cases: Coupled harmonic oscillators and Hooke's atom}
\label{sec:MultiCases}

\begin{figure}[t]
    \includegraphics[scale=0.55]{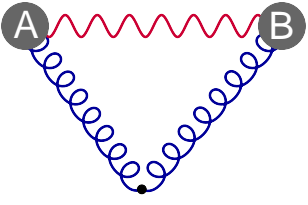}
    \caption{(color online) A set of two coupled harmonic oscillators. The binding potential (blue line) is harmonic, while the A-B coupling (red line) can be differently defined. 
	}
	\label{fig:HookeAtom}
\end{figure}

In this section we analyze two-photon scattering off multi-particle targets. For this purpose, we consider the target being a set of two coupled harmonic oscillators, as displayed in Fig. \ref{fig:HookeAtom}. The hamiltonian of such a system is:
\begin{align}
\opscalar H_0&= \sum_{i=A,B}\left(\frac{\opvector p_i^2}{2m}+\frac{m\omega^2}{2}\opvector r^2_i\right)
+
V(\opvector r_A-\opvector r_B)~,
\end{align}
where $V(\opvector r_A-\opvector r_B)$ is $any$ coupling potential between the oscillators. Let us consider, for simplicity, an incident photon that is linearly polarized along the $x$-direction. As showed in Sec. \ref{sec:Cancellation}, we may use $\comm{V(\opvector r_A-\opvector r_B)}{\opscalar O_1}=0$. Therefore $
\comm{\opscalar H_0}{\opscalar O_1}=-i\hbar c m \omega^2 (x_A+x_B)
$ and
$\comm{\opscalar H_0}{\comm{\opscalar H_0}{\opscalar O_1}}=\hbar^2\omega^2\opscalar O_1$. As a consequence, equations \eqref{eq:TinHO} and \eqref{eq:TfinalUND} hold also in this multi-particle case. 
Then, similarly to Eq. \eqref{eq:Tfinal}, the transition operator $\opscalar T$ turns out to be
\begin{align}
\opscalar{T}&=-i\frac{\hbar^2\omega^2}{E_1^2-\hbar^2\omega^2}\sqrt{\frac{2Mc^2}{\hbar\omega}}\opscalar{A}^\dagger_{1x, \gamma}~,
\label{eq:TfinalMulti}
\end{align}
where $\opscalar{A}^\dagger_{1x,\gamma}=\sqrt{\frac{M\omega}{2\hbar}}\left(
\opscalar R_x-\frac{i}{M\omega}\gamma \opscalar P_x
\right)$, and $M=2m$. $\opscalar R_x$ and $\opscalar P_x$ are the projections along the $x$-axis of the operators related to the center-of-mass coordinates; these are defined as $\opvector R = (\opvector r_A+\opvector r_B)/2$ and $\opvector P= \opvector p_{A}+ \opvector p_{B}$.
%

To further proceed, one needs to define the interaction that yields the coupling potential $V(\opvector r_A-\opvector r_B)$, so to define the states $\ket{i}$ and $\ket{f}$. As first example, let us choose the coupling potential to be harmonic. Harmonic coupling is mostly considered in crystals (classic phonon theory) \cite{tsirelson1996electron}, but also in many other physics fields, such as astrophysics \cite{srednicki1993entropy} or quantum many-body systems \cite{saito2007fluctuation}. With this choice, the target eigenstates $\ket{\Psi}$ are factorized into the three cartesian coordinates,
$
\ket{\Psi}=\ket{\phi_x}\ket{\phi_y}\ket{\phi_z}
$, 
where the vector state for each axis is itself factorized into center-of-mass and relative (inter-particle) coordinates as \cite{jonker2016entanglement}
\begin{equation}
\ket{\phi_j}=\frac{\big(\opscalar A_{1j}^\dagger\big)^{n_{1j}}\big(\opscalar A_{2j}^\dagger\big)^{n_{2j}}}{\sqrt{n_{1j}!n_{2j}!}}
\ket{00}
\end{equation}
for any $j=x,y,z$, while $n_{ij}=0,1,2,...$ is the excitation number. The ladder operator related to the center-of-mass coordinates is defined as $\opscalar{A}^\dagger_{1j}=\sqrt{\frac{M\omega}{2\hbar}}\left(
\opscalar R_j-\frac{i}{M\omega} \opscalar P_j
\right)=\opscalar{A}^\dagger_{1j,\gamma=1}$. On the other hand, $\opscalar{A}^\dagger_{2j}$ is the ladder operator related to the relative coordinates, and is not important for our analysis. 
Since the transition operator \eqref{eq:TfinalMulti} does not contain operators related to relative coordinates, it follows that 
the Rayleigh scattering amplitude is the same as in Eq. \eqref{eq:A12HO}
with $a_j\to \opscalar{A}_{1j}$ and $a_{j,\gamma}\to \opscalar{A}_{1j,\gamma}$.
The final form of the total amplitude is thus the one showed in Eq. \eqref{eq:FinalMHO} and displayed in Fig. \ref{fig:HORay}. In conclusion, although the target bound structure in this multi-particle case is richer than in the single-particle case, the Rayleigh scattering amplitudes are the same, due to the fact that the transition operator only contains operators related to the center-of-mass coordinates, and that the wavefunction is factorized into center-of-mass and relative coordinates. 

As a second example, we consider the coupling potential to be of Coulomb type: 
$V(\opvector r_A-\opvector r_B)=\alpha/|\opvector r_A - \opvector r_B|$, where $\alpha$ is the coupling constant. With this choice, the set of coupled harmonic oscillators in Fig. \ref{fig:HookeAtom} is known as Hooke's atom \cite{o2003wave}. The Hooke's atom is an atomic model for helium that approximates the Coulomb interaction between atomic electrons and nucleus with a harmonic interaction, while retaining the full electron-electron repulsion term in the hamiltonian. For this reason, it is considered important in quantum chemistry and physics for the study of electron-electron correlations and quantum entanglement \cite{gori2009study, manzano2010quantum}. The ground state wavefunction of the Hooke's atom is known analytically for many values of $\omega$. Such a wavefunction can be written in a factorized form as $\Psi(\bm R, \bm u)=\chi(\bm R)\, \phi(\bm u)$, where $\bm u=\bm r_1-\bm r_2$ is the relative coordinate and $\bm R$ (defined above) is the center-of-mass coordinate. While the function $\phi(\bm u)$ is specific to the Hooke's hamiltonian, the function $\chi(\bm R)$ turns out to be the wavefunction of a quantum harmonic oscillator. $\chi(\bm R)$ describes the movement of the center of mass of the coupled harmonic oscillators in Fig. \ref{fig:HookeAtom}. Let us now come back to the calculation of the Rayleigh scattering amplitude off the Hooke's atom. Since the transition operator \eqref{eq:TfinalMulti} does not contain operators related to relative coordinates, it will only act on the wavefunction $\chi(\bm R)$, and will consequently lead, once again, to the same transition amplitude we calculated for the single-particle harmonic oscillator in Sec. \ref{sec:sampleH}. 

The calculations above lead us to another interesting result, which can be formulated as follows: Irrespectively of the choice of the coupling potential, the Rayleigh scattering off coupled- and uncoupled- harmonic oscillators is characterized by the same transition amplitude, as long as the wavefunction can be separated into center-of-mass and relative coordinates. In other words, the Rayleigh scattering is not sensible to the inter-particle coupling potential, provided that the mentioned hypotheses are satisfied. The same result would be obtained for $N$ coupled harmonic oscillators.

\section{Further studies and extensions}
\label{sec:Ext}
Similarly to what showed in Secs. \ref{sec:SimpleCases} and \ref{sec:MultiCases}, one could compute light scattering (as well as any two-photon processes) off other kinds of potential. The computation of the transition amplitude could be algebraically more complicated but certainly feasible. In other words, with the formalism developed in this paper one can calculate (dipole) two-photon transitions irrespectively of the complexity of the potential. If a closed form solution for the transition amplitude cannot be found, a solution given by convergence can be sought, provided that a) the potential is not singular anywhere in the Real space, and that b) the photons have sufficiently high energy. 

The present study can be extended to higher multipoles. In order to consider all multipoles one needs to replace $\uvector \epsilon_{1,2} \cdot \opvector p \to \uvector\epsilon_{1,2} \cdot \opvector p e^{\pm i\bm k_{1,2} \opvector{r}/\hbar}$ in Eq. \eqref{eq:onephampl}, as well as in $\opscalar O_1$ in Eq. \eqref{eq:commRule}, where $k_{1,2}$ are the photon momenta. Furthermore, one needs to adjust also the classical term \eqref{eq:0thorder}, so to replace the Thomson term with the so-called Form Factor term \cite{ssa2015}. The resulting computation for the matrix amplitude would be more difficult but certainly feasible, at least if one looks for convergence at high photon energies. A closed form solution might be in fact not be available when high multipoles are considered, even for the simpler cases described in Secs. \ref{sec:SimpleCases} and \ref{sec:MultiCases}. 

An extension to relativistic quantum mechanics is analogously possible since the relativistic two-photon transition amplitude has a structure similar to the non-relativistic one. Even so, the commutators would be more challenging to compute, since the Dirac matrices do not commute with each other.

We so far considered the interaction potential to be that one given by (non-relativistic) quantum electrodynamics. 
If the interaction potential were different, equations \eqref{eq:exp} and \eqref{eq:commRule} would still hold, as long as second order perturbation theory can be applied. The whole formalism here developed would be therefore unchanged. What would need an amendment is the definition of the operator $\opscalar O_n$, of equation \eqref{eq:onephampl}, and of the classical term \eqref{eq:0thorder}. The amendment would be the replacement of the potential operators. 

\section{Summary and conclusions}
\label{sec:Concl}
We developed a new method for evaluating two-photon processes that replaces the (well-know) summation over the intermediate states by a series of commutator operators. By focusing on Raman and Rayleigh scattering as examples of two-photon processes, we showed how this method gives a clear distinction between the Thomson scattering, regarded as the classical term, and the next terms of powers of $\hbar$, regarded as quantum contributions. We applied this new method to study light scattering off several target hamiltonians, and we thereby obtained closed form solutions of the commutator series for the simpler potential cases, while we looked for convergence in the case of more complex potentials.

In the course of our analysis, we derived several results. First, we derived a general correction to the Thomson approximation, as well as an energy-dependence law for the cross section, which is valid for any target potential within the dipole approximation. Furthermore, we found an analytical transformation from the ground state to the perturbed state of a harmonic oscillator immersed into radiation.
We also showed that quantum contributions are vanishing (or significantly reduced) for targets composed by identical particles that are interacting with light. Moreover, we showed that linearly polarized light only probes the target binding potential along the polarization axis, under the assumption that such a potential is decoupled in cartesian coordinates. Finally, we demonstrated that as long as a) the target binding potential can be approximated to harmonic,  and b) the target wavefunction can be separated into relative and center-of-mass coordinates, then the elastic scattering amplitude is independent of inter-particle potentials.

As mentioned in previous sections, two-photon processes are applied in many scientific areas. The present work can be therefore potentially useful for forthcoming studies - in quantum chemistry, biology, crystals, mesoscopic systems, many-body physics, quantum optics and fundamental physics - that aim at analyzing two-photon processes beyond the Thomson or single resonance approximations \cite{npg2007}, for which the infinite summation over the target intermediate states is difficult or unfeasible to calculate, or where a clear distinction between classical and quantum contributions is sought.

\section{Acknowledgments}
P.~A. acknowledges the support of the FCT, under Contracts No. \emph{SFRH/BPD/92329/2013}.
This work was funded by the Portuguese Funda\c{c}\~{a}o para a Ci\^{e}ncia e a Tecnologia (FCT/MCTES/PIDDAC) under grant UID/FIS/04559/2013 (LIBPhys).\\
L. S. acknowledges financial support from the People Programme (Marie Curie Actions) of the European Union's Seventh Framework Programme (FP7/2007-2013 ) under REA Grant Agreement No. [291734].\\
F. F acknowledges partial support from the University of Applied Sciences BFI, Vienna.

\section{APPENDIX}
Here we explicitly write the scattering operator $\opscalar O$ for the case investigated in Sec. \ref{sec:Morse}, that is Rayleigh scattering off a particle vibrating along the $x$-axis, the vibration being modeled by a Morse potential. For simplicity let us set units such that $(V_0, a, x_0, m, \hbar, c)=(1,1/3,0,1,1,1)$. The ground state of the Morse potential is
\begin{equation}
\varphi_0=2^{\frac{9}{\sqrt{2}}-\frac{3}{4}} 3^{3 \sqrt{2}-1} e^{\frac{x}{3}-3 
\sqrt{2} e^{-x/3}} e^{-\frac{x}{3}(\frac{1}{2}+3 \sqrt{2})} 
\sqrt{\frac{6 \sqrt{2}-1}{\Gamma \left(6 \sqrt{2}\right)}}~.
\end{equation}
The component of the operator $\opscalar O$ up to the $6^{th}$ order are found to be as follows:
\begin{widetext}
\begin{align*}
\opscalar{O}_1=&- \opscalar{p}_x~,\qquad
\opscalar{O}_2= \frac{2}{3} i e^{-2 \opscalar{x}/3}-\frac{2}{3} i e^{-\opscalar{x}/3} ~,\qquad
\opscalar{O}_3=  -\frac{2}{9} \left(\opscalar{p}_xe^{-2 \opscalar{x}/3}\right)+\frac{1}{9} \left(\opscalar{p}_xe^{-\opscalar{x}/3}\right)-\frac{2}{9} \left(e^{-2 \opscalar{x}/3}\opscalar{p}_x\right)+\frac{1}{9} \left(e^{-\opscalar{x}/3}\opscalar{p}_x\right)~,\\[0.4cm]
\opscalar{O}_4=& -\frac{2}{27} i \left(\opscalar{p}_x^2e^{-2 \opscalar{x}/3}\right)+\frac{1}{54} i \left(\opscalar{p}_x^2e^{-\opscalar{x}/3}\right)-\frac{2}{27} i \left(e^{-2 \opscalar{x}/3}\opscalar{p}_x^2\right)+\frac{1}{54} i \left(e^{-\opscalar{x}/3}\opscalar{p}_x^2\right)-\frac{4}{27} i \left(\opscalar{p}_xe^{-2 \opscalar{x}/3}\opscalar{p}_x\right)+\frac{1}{27} i \left(\opscalar{p}_xe^{-\opscalar{x}/3}\opscalar{p}_x\right)\\[0.4cm]
&-\frac{2}{9} i \left(e^{-\opscalar{x}/3}e^{-2 \opscalar{x}/3}\right)-\frac{2}{9} i \left(e^{-2 \opscalar{x}/3}e^{-\opscalar{x}/3}\right)+\frac{8}{27} i e^{-4 \opscalar{x}/3}+\frac{4}{27} i e^{-2 \opscalar{x}/3}~,\\[0.4cm]
\opscalar{O}_5  = & \frac{2}{81} \left(\opscalar{p}_x^3e^{-2 \opscalar{x}/3}\right)-\frac{1}{324} 
\left(\opscalar{p}_x^3e^{-\opscalar{x}/3}\right)+\frac{2}{81} \left(e^{-2 \opscalar{x}/3}\opscalar{p}_x^3\right)-\frac{1}{324} \left(e^{-\opscalar{x}/3}\opscalar{p}_x^3\right)+\frac{2}{27} 
\left(\opscalar{p}_x^2e^{-2 \opscalar{x}/3}\opscalar{p}_x\right)-\frac{1}{108} \left(\opscalar{p}_x^2e^{-\opscalar{x}/3}\opscalar{p}_x\right)\\[0.4cm]
&+\frac{2}{27} \left(\opscalar{p}_xe^{-2 \opscalar{x}/3}\opscalar{p}_x^2\right)-\frac{1}{108} 
\left(\opscalar{p}_xe^{-\opscalar{x}/3}\opscalar{p}_x^2\right)-\frac{28}{81} \left(\opscalar{p}_xe^{-4 \opscalar{x}/3}\right)-\frac{7}{81} \left(\opscalar{p}_xe^{-2 \opscalar{x}/3}\right)-\frac{28}{81} \left(e^{-4 \opscalar{x}/3}\opscalar{p}_x\right)-\frac{7}{81} \left(e^{-2 \opscalar{x}/3}\opscalar{p}_x\right)\\[0.4cm]
&+\frac{1}{9} 
\left(\opscalar{p}_xe^{-\opscalar{x}/3}e^{-2 \opscalar{x}/3}\right)-\frac{8}{81} \left(e^{-2 \opscalar{x}/3}\opscalar{p}_xe^{-2 
\opscalar{x}/3}\right)+\frac{14}{81} \left(e^{-\opscalar{x}/3}\opscalar{p}_xe^{-2 \opscalar{x}/3}\right)+\frac{5}{27} \left(\opscalar{p}_xe^{-2 \opscalar{x}/3}e^{-\opscalar{x}/3}\right)+\frac{14}{81} 
\left(e^{-2 \opscalar{x}/3}\opscalar{p}_xe^{-\opscalar{x}/3}\right)\\[0.4cm]
&-\frac{2}{81} \left(e^{-\opscalar{x}/3}\opscalar{p}_xe^{-\opscalar{x}/3}
\right)+\frac{5}{27} \left(e^{-\opscalar{x}/3}e^{-2 \opscalar{x}/3}\opscalar{p}_x\right)+\frac{1}{9} 
\left(e^{-2 \opscalar{x}/3}e^{-\opscalar{x}/3}\opscalar{p}_x\right)~,\\[0.4cm]
\opscalar{O}_6  = &\frac{56}{243} i \left(e^{-2 \opscalar{x}/3}e^{-4 \opscalar{x}/3}\right)+\frac{40}{81} i 
e^{-4 \opscalar{x}/3}-\frac{38}{81} i \left(e^{-\opscalar{x}/3}e^{-4 \opscalar{x}/3}\right)-\frac{8}{27} i \left(\opscalar{p}_x^2e^{-4 \opscalar{x}/3}\right)+\frac{8}{27} i 
\left(e^{-4 \opscalar{x}/3}e^{-2 \opscalar{x}/3}\right)-\frac{14}{243} i 
\left(e^{-\opscalar{x}/3}e^{-2 \opscalar{x}/3}\right)\\[0.4cm]
&-\frac{1}{27} i \left(\opscalar{p}_x^2e^{-2 \opscalar{x}/3}\right)+\frac{2}{243} i \left(\opscalar{p}_x^4e^{-2 \opscalar{x}/3}\right)-\frac{38}{81} i 
\left(e^{-4 \opscalar{x}/3}e^{-\opscalar{x}/3}\right)-\frac{2}{27} i \left(e^{-2 
\opscalar{x}/3}e^{-\opscalar{x}/3}\right)-\frac{i \left(\opscalar{p}_x^4e^{-\opscalar{x}/3}\right)}{1944}-\frac{8}{27} i \left(e^{-4 \opscalar{x}/3}\opscalar{p}_x^2\right)\\[0.4cm]
&-\frac{1}{27} i 
\left(e^{-2 \opscalar{x}/3}\opscalar{p}_x^2\right)+\frac{2}{243} i \left(e^{-2 \opscalar{x}/3}\opscalar{p}_x^4\right)-\frac{i \left(e^{-\opscalar{x}/3}\opscalar{p}_x^4\right)}{1944}-\frac{52}{243} i 
\left(e^{-2 \opscalar{x}/3}e^{-\opscalar{x}/3}e^{-2 \opscalar{x}/3}\right)+\frac{10}{243} i 
\left(\opscalar{p}_x^2e^{-\opscalar{x}/3}e^{-2 \opscalar{x}/3}\right)\\[0.4cm]
&-\frac{8}{81} i \left(e^{-2 
\opscalar{x}/3}\opscalar{p}_x^2e^{-2 \opscalar{x}/3}\right)+\frac{17}{162} i \left(e^{-\opscalar{x}/3}\opscalar{p}_x^2e^{-2 \opscalar{x}/3}
\right)+\frac{64}{243} i \left(e^{-\opscalar{x}/3}e^{-2 \opscalar{x}/3}e^{-\opscalar{x}/3}\right)+\frac{55}{486} i \left(\opscalar{p}_x^2e^{-2 \opscalar{x}/3}e^{-\opscalar{x}/3}\right)\\[0.4cm]
&+\frac{17}{162} i \left(e^{-2 \opscalar{x}/3}\opscalar{p}_x^2e^{-\opscalar{x}/3}\right)-\frac{1}{81} i \left(e^{-\opscalar{x}/3}\opscalar{p}_x^2e^{-\opscalar{x}/3}\right)-\frac{136}{243} 
i \left(\opscalar{p}_xe^{-4 \opscalar{x}/3}\opscalar{p}_x\right)-\frac{17}{243} i \left(\opscalar{p}_xe^{-2 \opscalar{x}/3}\opscalar{p}_x\right)+\frac{8}{243} i \left(\opscalar{p}_x^3e^{-2 \opscalar{x}/3}\opscalar{p}_x\right)\\[0.4cm]
&-\frac{1}{486} i 
\left(\opscalar{p}_x^3e^{-\opscalar{x}/3}\opscalar{p}_x\right)+\frac{55}{486} i \left(e^{-\opscalar{x}/3}e^{-2 
\opscalar{x}/3}\opscalar{p}_x^2\right)+\frac{4}{81} i \left(\opscalar{p}_x^2e^{-2 \opscalar{x}/3}\opscalar{p}_x^2\right)+\frac{10}{243} i \left(e^{-2 \opscalar{x}/3}e^{-\opscalar{x}/3}\opscalar{p}_x^2\right)-\frac{1}{324} i \left(\opscalar{p}_x^2e^{-\opscalar{x}/3}\opscalar{p}_x^2\right)+\\[0.4cm]
&\frac{8}{243} i 
\left(\opscalar{p}_xe^{-2 \opscalar{x}/3}\opscalar{p}_x^3\right)-\frac{1}{486} i \left(\opscalar{p}_xe^{-\opscalar{x}/3}\opscalar{p}_x^3\right)-\frac{8}{81} i \left(\opscalar{p}_xe^{-2 \opscalar{x}/3}\opscalar{p}_xe^{-2 \opscalar{x}/3}\right)+\frac{26}{243} i \left(\opscalar{p}_xe^{-\opscalar{x}/3}\opscalar{p}_xe^{-2 \opscalar{x}/3}\right)\\[0.4cm]
&+\frac{49}{243} i \left(\opscalar{p}_xe^{-2 \opscalar{x}/3}\opscalar{p}_xe^{-\opscalar{x}/3}\right)-\frac{1}{81} 
i \left(\opscalar{p}_xe^{-\opscalar{x}/3}\opscalar{p}_xe^{-\opscalar{x}/3}\right)+\frac{10}{81} i 
\left(\opscalar{p}_xe^{-\opscalar{x}/3}e^{-2 \opscalar{x}/3}\opscalar{p}_x\right)-\frac{8}{81} i \left(e^{-2 
\opscalar{x}/3}\opscalar{p}_xe^{-2 \opscalar{x}/3}\opscalar{p}_x\right)\\[0.4cm]
&+\frac{49}{243} i \left(e^{-\opscalar{x}/3}\opscalar{p}_xe^{-2 \opscalar{x}/3}\opscalar{p}_x\right)+\frac{10}{81} i \left(\opscalar{p}_xe^{-2 \opscalar{x}/3}e^{-\opscalar{x}/3}\opscalar{p}_x\right)+\frac{26}{243} i \left(e^{-2 \opscalar{x}/3}\opscalar{p}_xe^{-\opscalar{x}/3}\opscalar{p}_x\right)-\frac{1}{81} 
i \left(e^{-\opscalar{x}/3}\opscalar{p}_xe^{-\opscalar{x}/3}\opscalar{p}_x\right)~.
\end{align*}
To compute the above elements we used the $Quantum$ Mathematica package \cite{gom2013}.
\end{widetext}


\bibliography{comu_articles}

\end{document}